\documentclass{webofc}
\usepackage[varg]{txfonts}   

\usepackage{multirow}

\newcommand{\fref}[1]{Fig. \ref{#1}}
\newcommand{\tref}[1]{Tab.~\ref{#1}}

\begin{document}

\title{Quenched glueball spectrum from functional equations}

\author{\firstname{Markus Q.} \lastname{Huber}\inst{1}\fnsep\thanks{\email{markus.huber@physik.jlug.de}} \and
        \firstname{Christian S.} \lastname{Fischer}\inst{1,2}\fnsep\thanks{\email{christian.fischer@theo.physik.uni-giessen.de}} \and
        \firstname{H\`elios} \lastname{Sanchis-Alepuz}\inst{3}\fnsep\thanks{\email{helios.sanchis-alepuz@silicon-austria.com}}
}

\institute{
Institut f\"ur Theoretische Physik, Justus-Liebig-Universit\"at Giessen, Heinrich-Buff-Ring 16, 35392 Giessen, Germany
\and
Helmholtz Forschungsakademie Hessen f\"ur FAIR (HFHF), GSI Helmholtzzentrum f\"ur Schwerionenforschung, Campus Gie{\ss}en, 35392 Gie{\ss}en, Germany
\and
Silicon Austria Labs GmbH, Inffeldgasse 33, 8010 Graz, Austria
          }

\abstract{
We give an overview of results for the quenched glueball spectrum from two-body bound state equations based on the 3PI effective action.
The setup, which uses self-consistently calculated two- and three-point functions as input, is completely self-contained and does not have any free parameters except for the coupling.
The results for $J^{\mathsf{PC}}=0^{\pm+},2^{\pm+},3^{\pm+},4^{\pm+}$ are in good agreement with recent lattice results where available.
For the pseudoscalar glueball, we present first results from a two-loop complete calculation, rendering also the bound state calculation fully self-consistent.
}

\maketitle

\section{Introduction}
\label{sec:introduction}

The spectrum of bound states in quantum chromodynamics (QCD) is both experimentally and theoretically not completely known yet.
Although measured and calculated for many decades by now, there are still some surprises, as, e.g., the experimentally established existence of pentaquarks or the appearance of the XYZ states, but also some remaining gaps that need to be filled.
Glueballs, bound states of gluons, belong to the latter class.
They can be investigated by ongoing and future experiments like ATLAS, BESIII, PANDA or Glue-X.

On the theory side, their masses were calculated with several methods, but a consensus only exists for the case of quenched QCD, viz., QCD with infinitely heavy quarks.
In particular, lattice results for the lightest glueballs have been the benchmark for over twenty years now \cite{Morningstar:1999rf}.
These results were confirmed and extended by subsequent lattice computations \cite{Chen:2005mg,Athenodorou:2020ani} and are also supported by other methods, e.g., \cite{Szczepaniak:1995cw,Szczepaniak:2003mr,Dudal:2010cd,Janowski:2011gt,Eshraim:2012jv,Dudal:2013wja,Huber:2020ngt,Huber:2021yfy}.
For an overview, we refer to the reviews \cite{Klempt:2007cp,Crede:2008vw,Mathieu:2008me,Ochs:2013gi,Llanes-Estrada:2021evz}.

When quarks are included dynamically, on the other hand, the situation is not as clear for lattice calculations.
The reasons are manifold, among them a poor signal-to-noise ratio, incomplete operator bases and a challenging continuum extrapolation \cite{Gregory:2012hu}.
Another new feature is that glueballs can mix with pure quark states which can have the same quantum numbers.
This also aggravates the interpretation of experimental results.
A recent analysis of radiative $J/\Psi$-decays of BESIII data, however, identified the scalar glueball \cite{Sarantsev:2021ein} with a mass very close to the prediction of quenched QCD.

Using functional methods, the challenges are completely different than for lattice methods.
A crucial point is the input in form of two- and three-point functions that is required.
Initially, models were used as in \cite{Meyers:2012ka,Sanchis-Alepuz:2015hma,Souza:2019ylx,Kaptari:2020qlt}.
In \cite{Huber:2020ngt}, we used for the first time self-consistent input calculated from Dyson-Schwinger equations \cite{Huber:2020keu}.
As it turned out, this was decisive to obtain a description of the pseudoscalar and the scalar glueballs at the same time which was not possible before \cite{Sanchis-Alepuz:2015hma}.
As an additional benefit, the complete calculation is now self-contained and does not depend on any model parameters.
In a subsequent calculation, we extended the setup to nonzero spin \cite{Huber:2021yfy}.
In addition, we present here first results for a fully self-consistent calculation of the pseudoscalar glueball by including also the two-loop diagrams.

In the following, we present the setup of our calculations in Sec.~\ref{sec:setup} and the results in Sec.~\ref{sec:results}.
We close with a short summary and outlook.

\section{Setup}
\label{sec:setup}

\begin{figure*}[tb]
	\includegraphics[width=0.98\textwidth]{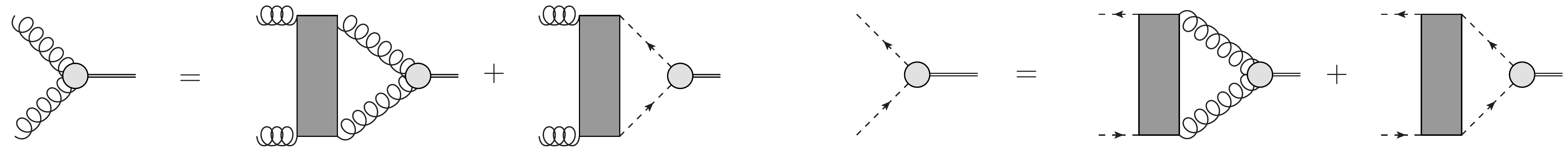}
	\caption{
		The coupled set of BSEs for a glueball made from two gluons and a pair of Faddeev-Popov (anti-)ghosts.
		Wiggly lines denote dressed gluon propagators, dashed lines denote dressed ghost propagators. 
		The gray boxes represent interaction kernels given in Fig.~\ref{fig:kernels}. The Bethe-Salpeter amplitudes of the glueball
		are denoted by gray disks. \label{fig:bses}
	}
	\vspace*{6mm}
	\includegraphics[width=0.48\textwidth]{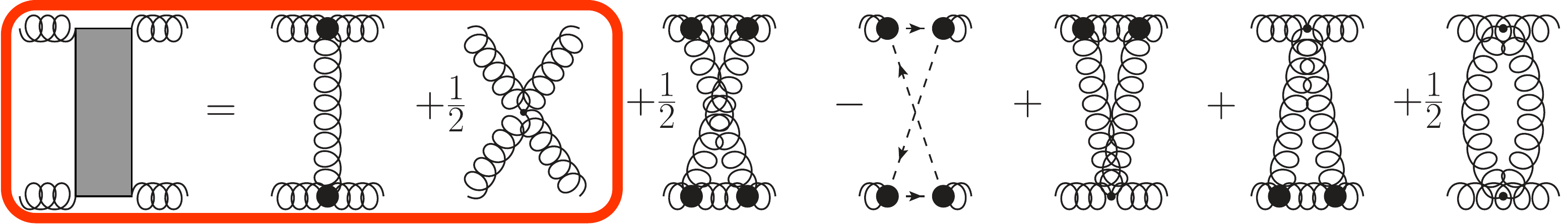}\\ 
	\vskip4mm
	\includegraphics[height=1.2cm]{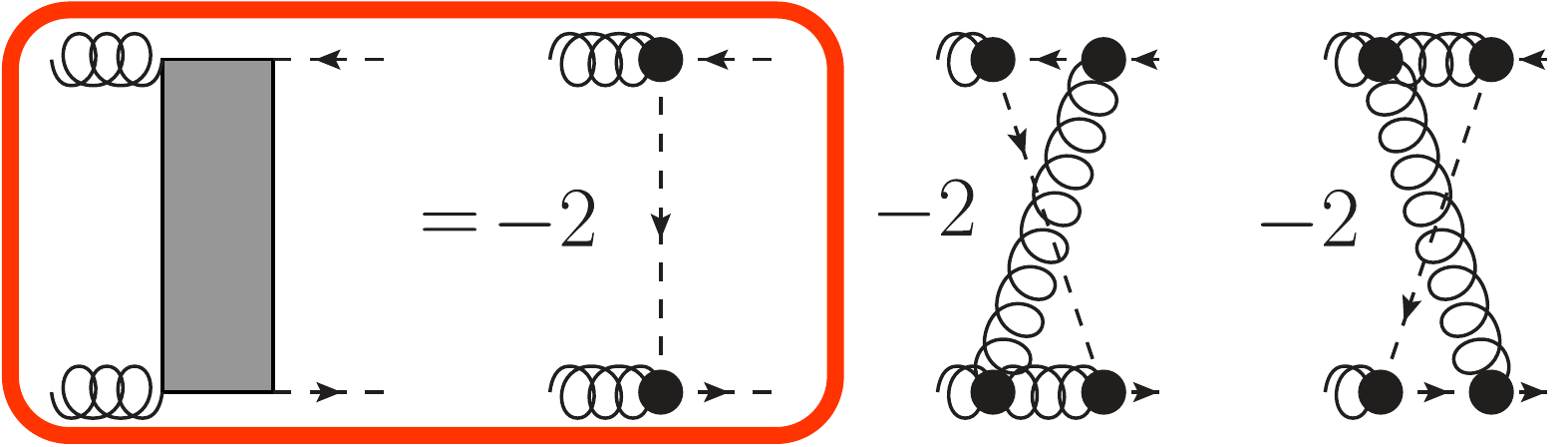}\hfill
	\includegraphics[height=1.2cm]{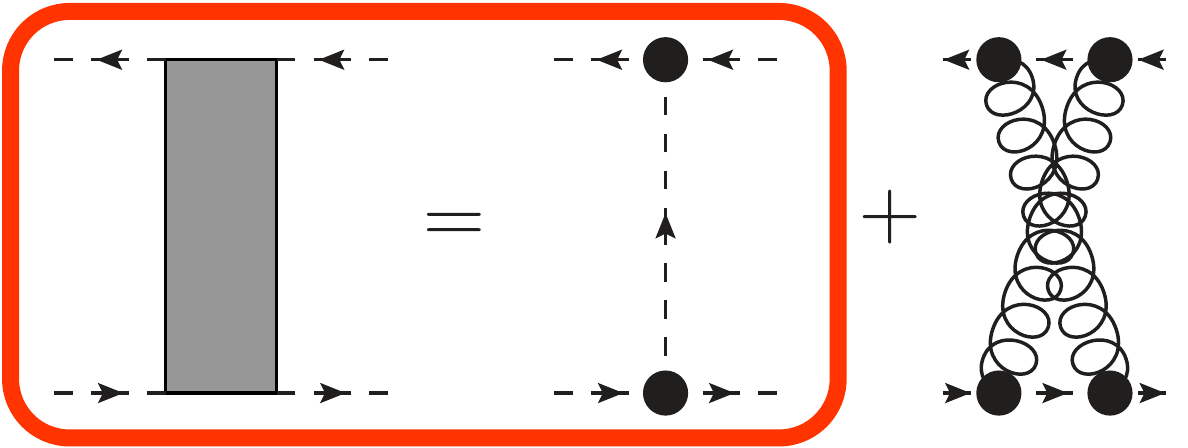}\hfill
	\includegraphics[height=1.2cm]{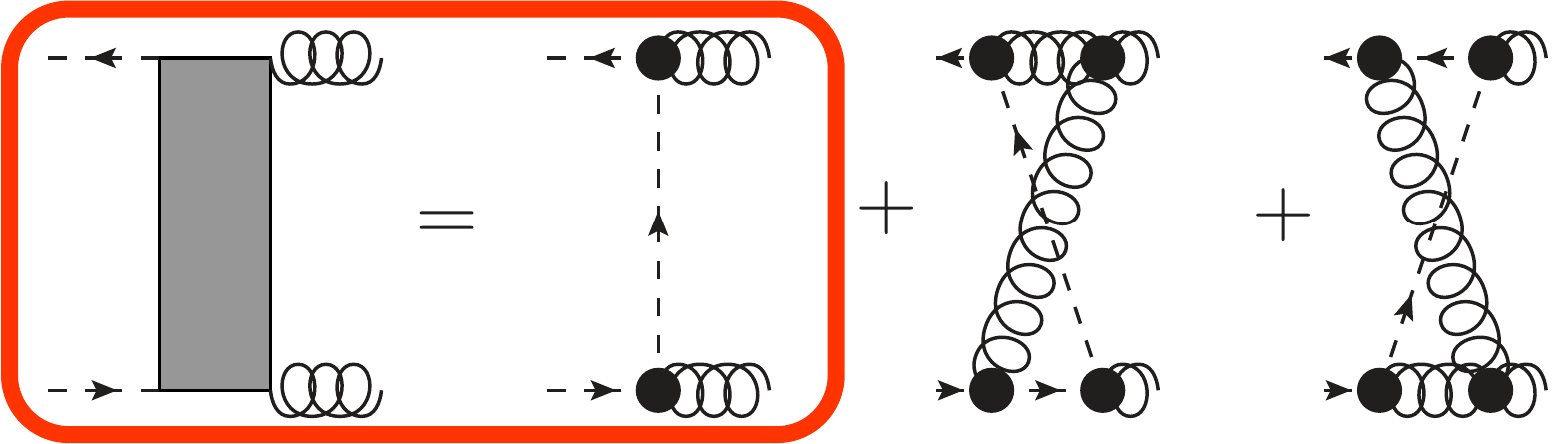}
	\caption{
		Interaction kernels from the three-loop 3PI effective action.
		All propagators are dressed; black disks represent dressed vertices.
		In our calculation, we include the diagrams inside the red rectangles. \label{fig:kernels}
	}
\end{figure*}

Glueballs can be described by gauge-invariant operators.
For example, $O(x)=F_{\mu\nu}(x) F^{\mu\nu}(x)$ has the quantum numbers of the scalar glueball and its mass can be extracted from the pole of the corresponding two-point function $\langle 0| O(x) O(y) | 0 \rangle$.
As opposed to the case of typically used mesonic operators, this function consists of parts with two, three and four gluons.
The calculation of the complete two-point function is a complicated task.
However, it can be sufficient to consider single pieces for the determination of masses.
The reason is that if a pole present is in one piece, this corresponds to the mass of the bound state, as long as no cancellation between different pieces occurs.
This is the rationale why we can use a two-body bound state equation to determine the mass of a complicated object like a glueball.
More details are provided in \cite{Huber:2021yfy}.
The resulting Bethe-Salpeter equation (BSE) is given in \fref{fig:bses}.
In general, as we necessarily work in a gauge-fixed setting, the BSE also contains ghost fields and a corresponding amplitude which we call the ghostball-part.

For the calculation, the interaction kernels need to be specified.
In analogy to the equations used to calculate the input correlation functions, see below, we derived them from a three-loop truncated 3PI effective action.
This yields the diagrams shown in \fref{fig:kernels}.
However, initially we only take into account the diagrams leading to one-loop diagrams, the reason being that the two-loop diagrams are computationally much more expensive.
Due to the good agreement of our results with lattice results, it seems that the two-loop contributions are subleading.
For the pseudoscalar glueball, which is the computationally least expensive, we performed also a calculation with the two-loop diagrams included.
In that case, the truncation is fully self-consistent.
For the mass determination, the calculation of the two-loop diagrams was computed with lower precision than the one-loop part.
However, for single eigenvalues we increased the precision and checked that our conclusion on the effect of two-loop diagrams remains valid.

\begin{figure*}[t]
	\includegraphics[width=0.48\textwidth]{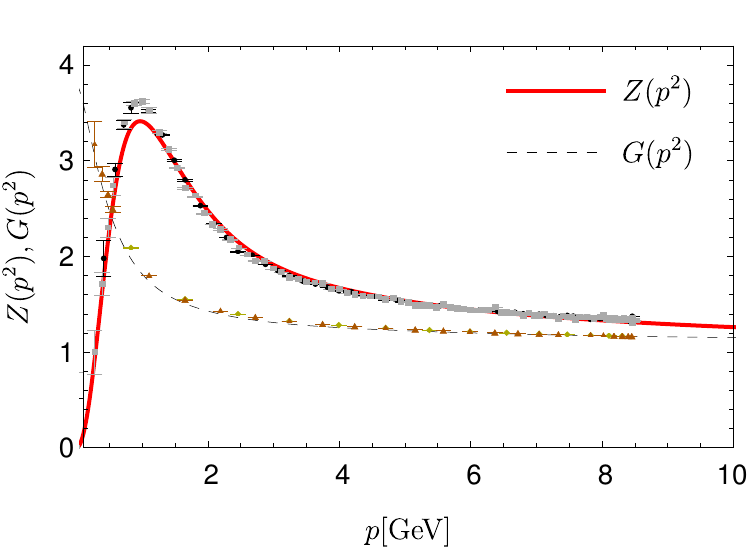}\hfill
	\includegraphics[width=0.48\textwidth]{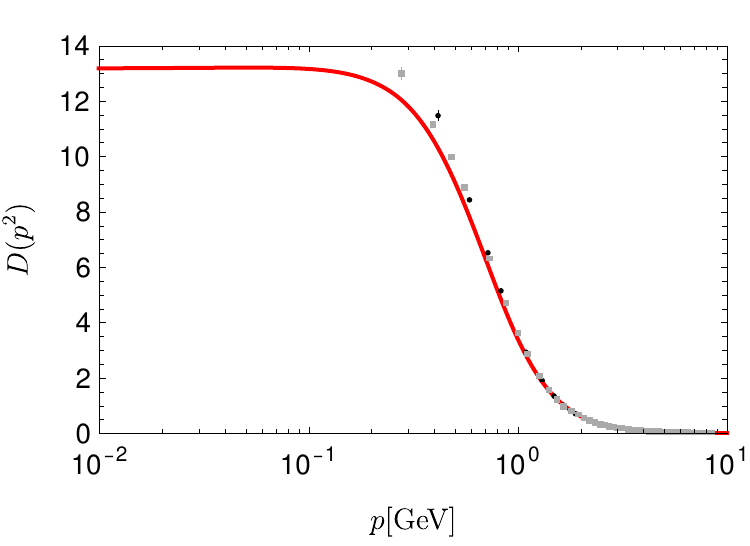}
	\caption{Gluon and ghost dressing functions $Z(p^2)$ and $G(p^2)$, respectively, (left) and gluon propagator $D(p^2)$ (right) in comparison to lattice data \cite{Sternbeck:2006rd}.
	}
	\label{fig:props}
	\includegraphics[width=0.48\textwidth]{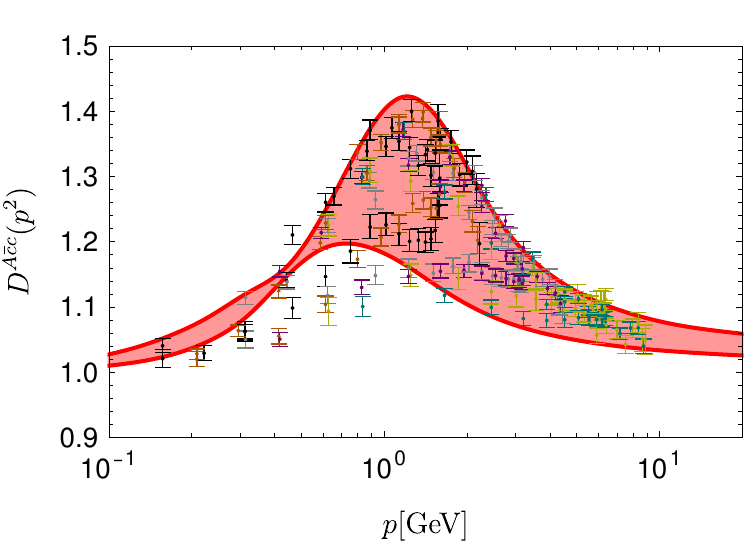}\hfill
	    \includegraphics[width=0.48\textwidth]{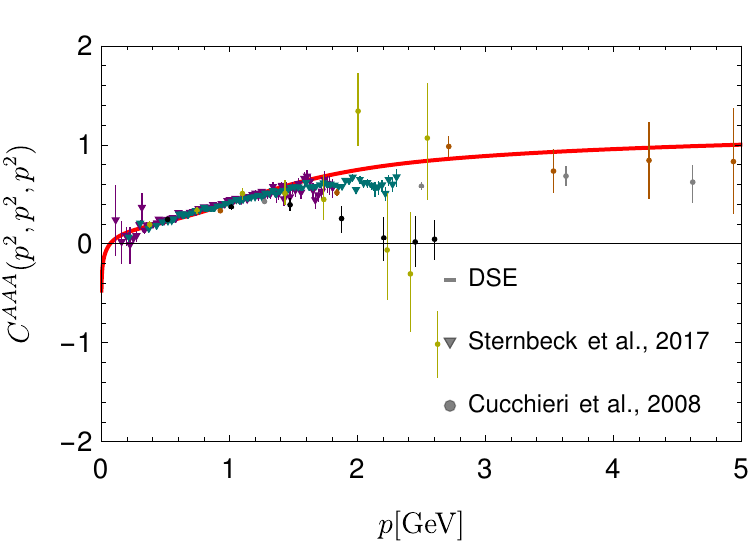}
	\caption{Left: Ghost-gluon vertex dressing function (full kinematic dependence) in comparison to $SU(2)$ lattice data \cite{Maas:2019ggf}.
		Right: Three-gluon vertex dressing function at the symmetric point in comparison to lattice data \cite{Cucchieri:2008qm,Sternbeck:2017ntv}), see Refs.~\cite{Athenodorou:2016oyh,Boucaud:2017obn} for similar results.
		}
    \label{fig:verts}
\end{figure*}

In this truncation, the BSE depends on the ghost and gluon propagators and the ghost-gluon and three-gluon vertices.
As mentioned above, we take them from a self-consistent, self-contained calculation \cite{Huber:2020keu}.
In Figs.~\ref{fig:props} and \ref{fig:verts}, we show a selection of the input in comparison with lattice results.
We want to stress that this corresponds to one of several possible solutions which vary in their infrared behavior, see, e.g., \cite{Boucaud:2008ji,Fischer:2008uz,Alkofer:2008jy,Maas:2009se,Maas:2011se,Sternbeck:2012mf,Huber:2018ned,Eichmann:2021zuv}.
For the spin-0 case, we tested that the masses agree for different solutions and we thus continued with only one solution for higher spins.

The BSE is solved as an eigenvalue equation.
By varying the total momentum $P$, curves for the eigenvalues $\lambda_i$ are determined.
Physical solutions correspond to eigenvalues $\lambda_i=1$ and the corresponding value of $P$ determines the mass: $M^2=-P^2$.
For time-like $P$, the propagators in the integrand are probed at complex values.
However, the input we use is only available for Euclidean momenta.
Corresponding calculations in the complex plane are currently only available for much simpler truncations and restricted to the propagators \cite{Fischer:2020xnb}.
Thus, we can calculate the eigenvalue curves only for space-like $P$.
From these results, we extrapolate to the time-like side using Schlessinger's continued fraction method \cite{Schlessinger:1968spm,Tripolt:2018xeo}.
We tested this method in a model calculation \cite{Huber:2020ngt} for which it turned out to be very reliable at least up to approximately $2\,\text{GeV}$.
Beyond that, errors start to increase and need to be interpreted carefully.
For our highest mass results, we assign some level of uncertainty as indicated in \tref{tab:masses} and \fref{fig:spectrum}.
Errors estimated from the extrapolation are obtained by averaging over extrapolations from different subsets of eigenvalues.
Details can be found in~\cite{Huber:2020ngt}.

To calculate the masses for different spins and parities, appropriate bases need to be constructed.
The basis for spin $J$ consists for the ghostball- and glueball-parts of all tensors with $J$ and $J+2$ Lorentz tensors, respectively, which fulfill certain constraints.
For the former, they need to be traceless, symmetric and transverse with respect to the total momentum.
This leaves only one tensor.
For the glueball-part, however, these constraints only apply to the $J$ Lorentz indices for spin, but there are two additional indices from the gluon legs.
This increases the number of tensors, but as an additional constraint we need to ensure that we have a basis that is transverse in the gluon legs due to the transversality of the gluon propagator in the Landau gauge employed here.
If we included nontransverse parts, this would lead to inconsistencies in the eigenvalue equation that distorts the results.
Finally, the parity of the tensors needs to be taken into account.
The remaining tensor of the ghostball-part has positive parity which means that there is no ghostball-part for negative parity.
Following these considerations, one ends up with at most five tensors.
This is a rather small basis, but the expressions become quite lengthy with increasing $J$ due to the symmetrization and the required tracelessness.
The detailed expressions for the bases are given in \cite{Huber:2021yfy}.

\section{Results}
\label{sec:results}

\begin{figure*}[tb]
	\begin{center}
	\includegraphics[width=0.68\textwidth]{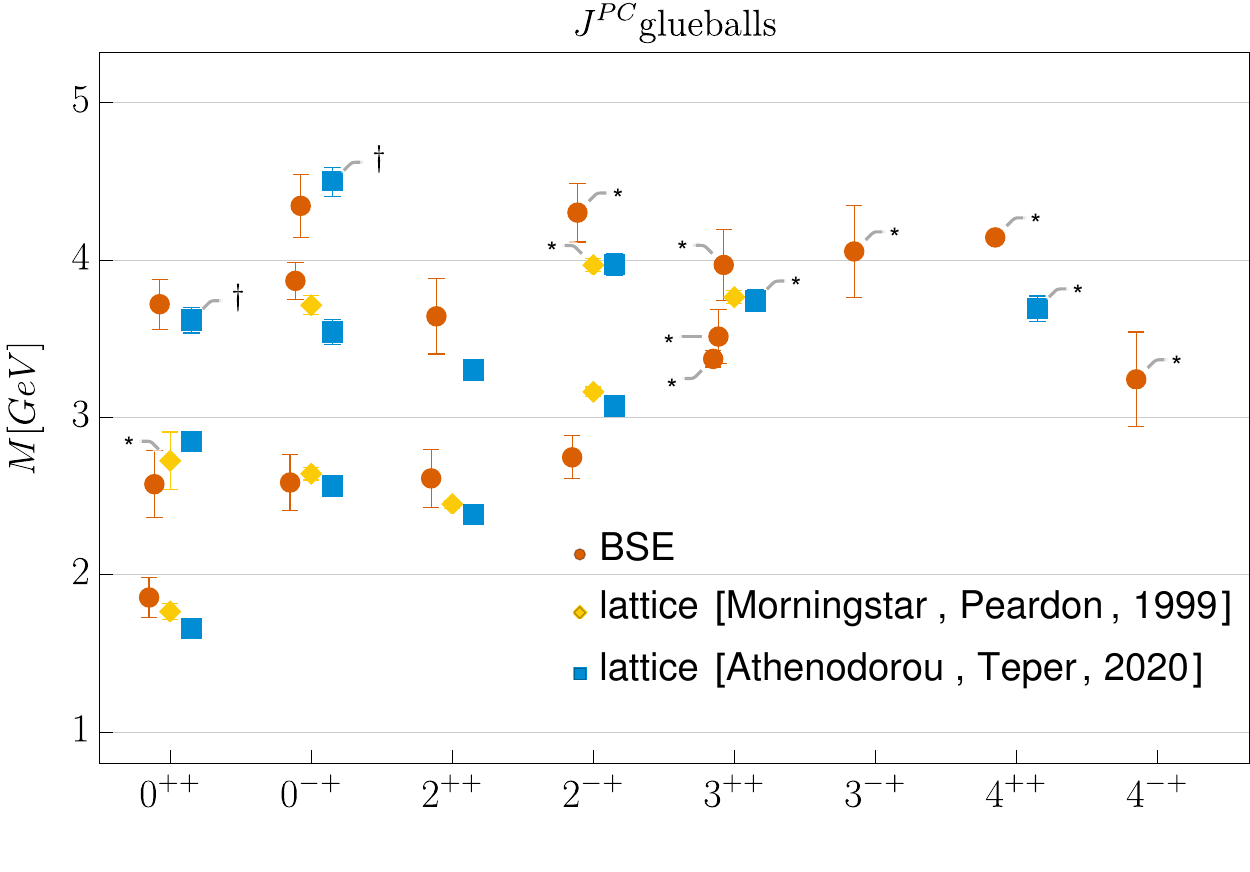}\\
	\includegraphics[width=0.68\textwidth]{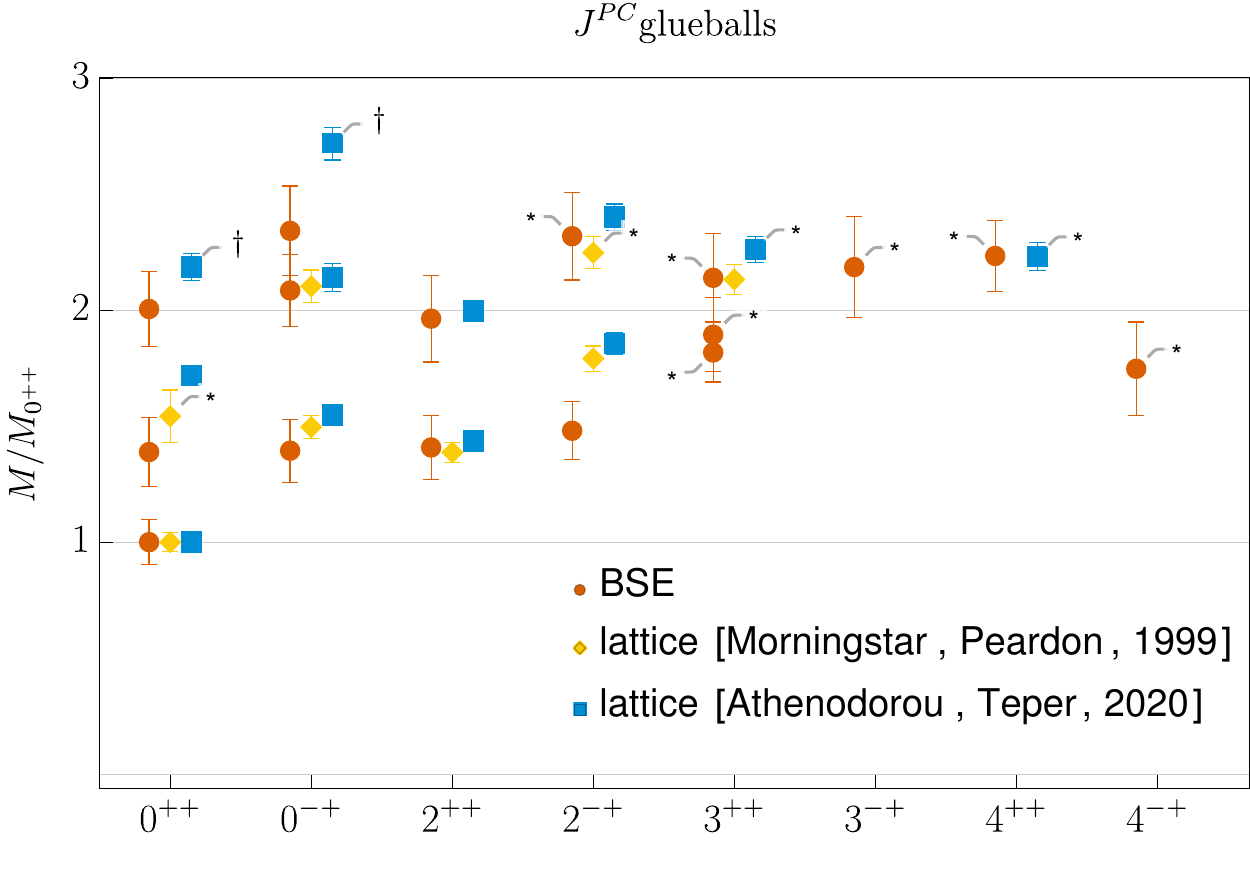}
	\end{center}
	\caption{
		Results for glueball ground states and excited states for the indicated quantum numbers from lattice simulations \cite{Morningstar:1999rf,Athenodorou:2020ani} and functional equations.
		In the upper plot, we display the glueball masses on an absolute scale set by $r_0=1/418(5)\,\text{MeV})$.
	    In the lower plot, we display the spectrum relative to the ground state.
	    Masses with $^\dagger$ are conjectured to be the second excited states.
		Masses with $^*$ come with some uncertainty in their identification in the lattice case or in the trustworthiness of the extrapolated value in the BSE case.
		}
	\label{fig:spectrum}
\end{figure*}

\begin{table*}[tb]
	\begin{center}
\caption{Ground and excited state masses $M$ of glueballs for various quantum numbers.
		Compared are lattice results from \cite{Morningstar:1999rf,Athenodorou:2020ani} with the results of this work and \cite{Huber:2020ngt}.
        For \cite{Morningstar:1999rf}, the errors are the combined errors from statistics and the use of an anisotropic lattices.
		For \cite{Athenodorou:2020ani}, the error is statistical only.
		In our results, the error comes from the extrapolation method and should be considered a lower bound on errors.
		All results use the same value for $r_0=1/(418(5)\,\text{MeV})$.
		The related error is not included in the table.
		Masses with $^\dagger$ are conjectured to be the second excited states.
		Masses with $^*$ come with some uncertainty in their identification in the lattice case or in the trustworthiness of the extrapolated value in the BSE case.
		}
		\begin{tabular}{|l||c|c|c|c|c|c|}
			\hline
			&  \multicolumn{2}{c|}{\cite{Morningstar:1999rf}} & \multicolumn{2}{c|}{\cite{Athenodorou:2020ani}} & \multicolumn{2}{c|}{This work}\\   
			\hline
			State &  $M\, [\text{MeV}]$& $M/M_{0^{++}}$ & $M\, [\text{MeV}]$& $M/M_{0^{++}}$ & $M\,[\text{MeV}]$ & $M/M_{0^{++}}$\\   
			\hline\hline
			$0^{++}$ & $1760 (50)$ & $1(0.04)$ & $1651(23)$ & $1(0.02)$ & $1850 (130)$ & $1(0.1)$\\
			\hline
			$0^{^*++}$ & $2720 (180)^*$ & $1.54(0.11)^*$ & $2840(40)$ & $1.72(0.034)$ & $2570 (210)$ & $1.39(0.15)$\\
			\hline
			\multirow{2}{*}{$0^{^{**}++}$} & \multirow{2}{*}{--} & \multirow{2}{*}{--} & $3650(60)^\dagger$ & $2.21(0.05)^\dagger$ & \multirow{2}{*}{$3720 (160)$} & \multirow{2}{*}{$2.01(0.16)$}\\
			& & & $3580(150)^\dagger$ & $2.17(0.1)^\dagger$ & &\\
			\hline
			$0^{-+}$ & $2640 (40) $ & $1.50(0.05)$ & $2600(40)$ & $1.574(0.032)$ & $2580 (180)$ & $1.39(0.14)$\\
			\hline
			$0^{^*-+}$ & $3710 (60)$ & $2.10(0.07)$ & $3540(80)$ & $2.14(0.06)$ & $3870 (120)$ & $2.09(0.16)$\\
			\hline
			\multirow{2}{*}{$0^{^{**}-+}$} & \multirow{2}{*}{--} & \multirow{2}{*}{--} & $4450(140)^\dagger$ & $2.7(0.09)^\dagger$ & \multirow{2}{*}{$4340 (200)$} & \multirow{2}{*}{$2.34(0.19)$}\\
			& & & $4540(120)^\dagger$ & $2.75(0.08)^\dagger$ & &\\
			\hline
			\hline
			$2^{++}$ & 2447(25) & 1.39(0.04) & 2376(32) & 1.439(0.028) & 2610(180) & 1.41(0.14)\\
			\hline
			$2^{^*++}$ & -- & -- & 3300(50) & 2(0.04) & 3640(240) & 1.96(0.19)\\
			\hline
			$2^{-+}$ & 3160(31) & 1.79(0.05) & 3070(60) & 1.86(0.04) & 2740(140) & 1.48(0.13)\\
			\hline
			$2^{^*-+}$ &  3970(40)$^*$ & 2.25(0.07)$^*$ & 3970(70) & 2.4(0.05) & 4300(190) & 2.32(0.19)\\
			\hline\hline
			$3^{++}$ & 3760(40) & 2.13(0.07) & 3740(70)$^*$ & 2.27(0.05)$^*$ & 3370(50)$^*$ & 1.82(0.13)$^*$\\
			\hline
			$3^{^*++}$ & -- & -- & -- & -- & 3510(170)$^*$ & 1.89(0.16)$^*$\\
			\hline
			$3^{^{**}++}$ & -- & -- & -- && 3970(220)$^*$ & 2.14(0.19)$^*$\\
			\hline
			$3^{-+}$ & -- & -- & -- & -- &  4050(290)$^*$ & 2.19(0.22)$^*$\\
			\hline
			\hline
			$4^{++}$ & -- & -- & 3690(80)$^*$ & 2.24(0.06)$^*$ & 4140(30)$^*$ & 2.23(0.15)$^*$\\
			\hline
			$4^{-+}$ & -- & -- & -- & -- & 3240(300)$^*$ & 1.75(0.2)$^*$\\
			\hline
		\end{tabular}
		\label{tab:masses}
	\end{center}
\end{table*}

We summarize the results for the glueball masses in \tref{tab:masses} and \fref{fig:spectrum}.
We also list results from lattice calculations.
To enable a meaningful comparison, we rescaled all results to a common value for the Sommer scale, $r_0=0.472(5)\,\text{fm}$, which was used in \cite{Athenodorou:2020ani}.
The original scale of the functional results was determined via the gluon propagator and a comparison with the lattice results of \cite{Sternbeck:2006rd}.
The indicated errors for the functional results are estimates from the extrapolation only and should be considered as a lower bound.
For higher spins, we reduced the numeric precision due to computation time.
From tests with low spin, we know that the ground state is not affected by that but excited states can vanish.
This is why we do not see excited states for all quantum numbers.
In general, the results are more uncertain for heavier masses due to the extrapolation deeper into the time-like region.
We indicate this by a star.
The results for $J^\mathsf{PC}=3^{++}$ and $4^{++}$ are somewhat exceptional, as the extrapolations lead to extremely small errors.
With regard to the stability of these states, we want make clear that the employed setup does not contain channels for decays.
Thus, all states are stable as long as this possibility is not included, e.g., via dedicated additional diagrams \cite{Fischer:2007ze}.
An interesting finding concerns spin one.
Often the Landau-Yang theorem is invoked to explain why spin-1 glueballs cannot be built from two gluons.
However, as explained in more detail in \cite{Huber:2021yfy}, in our approach it does not apply because the gluons are not on-shell.
Nevertheless, we did not find solutions for $J=1$.

For the pseudoscalar glueball we also performed a calculation with the two-loop diagrams included.
This glueball is technically the simplest one for two reasons: There is no ghostball-part and it has a one-dimensional basis.
Nevertheless, the calculation of the two-loop diagrams is very costly and we used only a low precision.
That calculation showed a very small change in the eigenvalues of less than 0.1\,\textperthousand.
More importantly, although in principle conceivable due to the sensitivity of the extrapolation to small changes in the input data, this did not lead to a measurable change in the mass.
Finally, we increased the precision further for a few eigenvalues and found that the agreement between the eigenvalues improves even further.
Thus we conclude that for the pseudoscalar glueball the two-loop diagrams are not only subleading but totally irrelevant.

\section{Summary and outlook}

We presented results for the glueball spectrum of pure Yang-Mills theory obtained in a functional framework with Bethe-Salpeter and Dyson-Schwinger equations.
The results are in good quantitative agreement with lattice results.
We also added some new states, e.g., second excited states of (pseudo)scalar glueballs.
The setup is completely free of any parameters to be tuned and thus our results constitute genuine predictions from a first principles calculation.

There are still some improvements that can be made.
One is the inclusion of two-loop diagrams which is necessary to confirm if they are indeed as subleading as the comparison with lattice results indicates.
This step is computationally demanding but conceptually straightforward.
We explored that for the technically simplest glueball, the pseudoscalar one, and found that the two-loop diagrams do not lead to any measurable effect on the mass or the amplitude.
This is the first case of a fully self-consistent calculation of a glueball from the three-loop truncated 3PI effective action.
More challenging is the circumvention of the extrapolation by working directly with time-like total momentum.
In terms of truncations, presently existing results \cite{Fischer:2020xnb} would need to be improved considerably as currently they are lacking two-loop diagrams in the propagators and dynamical vertices.
Beyond improvements and tests of the current setup, the inclusion of quarks constitutes the most interesting next step, as this would finally make contact to experiment.

\section*{Acknowledgments}

This work was supported by the DFG (German Research Foundation) grant FI 970/11-1 and by the BMBF under 
contracts No. 05P18RGFP1 and 05P21RGFP3.
This work has also been supported by Silicon Austria Labs (SAL), owned by the Republic of Austria, the Styrian Business Promotion Agency (SFG), the federal state of Carinthia, the Upper Austrian Research (UAR), and the Austrian Asso­ci­a­tion for the Elec­tric and Elec­tronics Industry (FEEI).

\bibliography{literature_glueballs_vConf21}

\end{document}